\documentclass [12pt] {article}
\usepackage{a4}

\usepackage{amsfonts}
\usepackage{amsmath}

  \def\pp{{\mathchoice
          {
              \kern 1pt%
              \raise 1pt
              \vbox{\hrule width5pt height0.4pt depth0pt
                    \kern -2pt
                    \hbox{\kern 2.3pt
                          \vrule width0.4pt height6pt depth0pt
                          }
                    \kern -2pt
                    \hrule width5pt height0.4pt depth0pt}%
                    \kern 1pt
           }
            {
              \kern 1pt%
              \raise 1pt
              \vbox{\hrule width4.3pt height0.4pt depth0pt
                    \kern -1.8pt
                    \hbox{\kern 1.95pt
                          \vrule width0.4pt height5.4pt depth0pt
                          }
                    \kern -1.8pt
                    \hrule width4.3pt height0.4pt depth0pt}%
                    \kern 1pt
            }
            {
              \kern 0.5pt%
              \raise 1pt
              \vbox{\hrule width4.0pt height0.3pt depth0pt
                    \kern -1.9pt  
                    \hbox{\kern 1.85pt
                          \vrule width0.3pt height5.7pt depth0pt
                          }
                    \kern -1.9pt
                    \hrule width4.0pt height0.3pt depth0pt}%
                    \kern 0.5pt
            }
            {
              \kern 0.5pt%
              \raise 1pt
              \vbox{\hrule width3.6pt height0.3pt depth0pt
                    \kern -1.5pt
                    \hbox{\kern 1.65pt
                          \vrule width0.3pt height4.5pt depth0pt
                          }
                    \kern -1.5pt
                    \hrule width3.6pt height0.3pt depth0pt}%
                    \kern 0.5pt
            }
        }}

  \def\mm{{\mathchoice
                       {
                             \kern 1pt
               \raise 1pt    \vbox{\hrule width5pt height0.4pt depth0pt
                                  \kern 2pt
                                  \hrule width5pt height0.4pt depth0pt}
                             \kern 1pt}
                       {
                            \kern 1pt
               \raise 1pt \vbox{\hrule width4.3pt height0.4pt depth0pt
                                  \kern 1.8pt
                                  \hrule width4.3pt height0.4pt depth0pt}
                             \kern 1pt}
                       {
                            \kern 0.5pt
               \raise 1pt
                            \vbox{\hrule width4.0pt height0.3pt depth0pt
                                  \kern 1.9pt
                                  \hrule width4.0pt height0.3pt depth0pt}
                            \kern 1pt}
                       {
                           \kern 0.5pt
             \raise 1pt  \vbox{\hrule width3.6pt height0.3pt depth0pt
                                  \kern 1.5pt
                                  \hrule width3.6pt height0.3pt depth0pt}
                           \kern 0.5pt}
                       }}

\catcode`@=11
\def\un#1{\relax\ifmmode\@@underline#1\else
        $\@@underline{\hbox{#1}}$\relax\fi}
\catcode`@=12

\let\du=\du                     

\def\a{\alpha}
\def\b{\beta}
\def\c{\chi}
\def\d{\delta}

\def\f{\phi}

\def\h{\eta}

\def\j{\psi}
\def\k{\kappa}
\def\l{\lambda}
\def\m{\mu}

\def\o{\omega}

\def\q{\theta}

\def\s{\sigma}

\def\z{\zeta}

\def\ve{\varepsilon}

\def\bo{{\raise-.5ex\hbox{\large$\Box$}}}               
\def\pa{\partial}                                       
\def\TH{{\raise.2ex\hbox{$\displaystyle \bigodot$}\mskip-4.7mu \llap H \;}}
\def\face{{\raise.2ex\hbox{$\displaystyle \bigodot$}\mskip-2.2mu \llap {$\ddot
        \smile$}}}                                      

\def\VEV#1{\left\langle #1\right\rangle}        
\def\abs#1{\left| #1\right|}                    
\def\leftrightarrowfill{$\mathsurround=0pt \mathord\leftarrow \mkern-6mu
        \cleaders\hbox{$\mkern-2mu \mathord- \mkern-2mu$}\hfill
        \mkern-6mu \mathord\rightarrow$}
\def\dvec#1{\vbox{\ialign{##\crcr
        \leftrightarrowfill\crcr\noalign{\kern-1pt\nointerlineskip}
        $\hfil\displaystyle{#1}\hfil$\crcr}}}           
\def\dt#1{{\buildrel {\hbox{\LARGE .}} \over {#1}}}     

\def\frac#1#2{{\textstyle{#1\over\vphantom2\smash{\raise.20ex
        \hbox{$\scriptstyle{#2}$}}}}}                   
\def\sfrac#1#2{{\vphantom1\smash{\lower.5ex\hbox{\small$#1$}}\over
        \vphantom1\smash{\raise.4ex\hbox{\small$#2$}}}} 
\def\bfrac#1#2{{\vphantom1\smash{\lower.5ex\hbox{$#1$}}\over
        \vphantom1\smash{\raise.3ex\hbox{$#2$}}}}       
\def\afrac#1#2{{\vphantom1\smash{\lower.5ex\hbox{$#1$}}\over#2}}    

\def\[{\lfloor{\hskip 0.35pt}\!\!\!\lceil}
\def\]{\rfloor{\hskip 0.35pt}\!\!\!\rceil}

\def\du#1#2{_{#1}{}^{#2}}

\def\un{\underline}
\def\fracmm#1#2{{{#1}\over{#2}}}

\def\low#1{{\raise -3pt\hbox{${\hskip 0.75pt}\!_{#1}$}}}

\def\Dot#1{\buildrel{_{_{\hskip 0.01in}\bullet}}\over{#1}}
\def\dt#1{\Dot{#1}}

\newskip\humongous \humongous=0pt plus 1000pt minus 1000pt
\def\caja{\mathsurround=0pt}
\def\eqalign#1{\,\vcenter{\openup2\jot \caja
        \ialign{\strut \hfil$\displaystyle{##}$&$
        \displaystyle{{}##}$\hfil\crcr#1\crcr}}\,}
\newif\ifdtup

\parskip=\medskipamount                 
\thicklines                         

\begin{document}
\thispagestyle{empty}

{\hbox to\hsize{
\vbox{\noindent October 2003 \hfill hep-th/0310152 }}}

\noindent
\vskip1.3cm
\begin{center}

{\Large\bf On the Universality of Goldstino Action~\footnote{Supported 
in part by the Japanese Society for Promotion of Science (JSPS) and the TMU 
fund}}
\vglue.2in

Tomoya Hatanaka~\footnote{Email address: thata@kiso.phys.metro-u.ac.jp}
 and Sergei V. Ketov~\footnote{Email address: ketov@phys.metro-u.ac.jp}

{\it Department of Physics, Faculty of Science\\
     Tokyo Metropolitan University\\
     1--1 Minami-osawa, Hachioji-shi\\
     Tokyo 192--0397, Japan}
\end{center}
\vglue.2in
\begin{center}
{\Large\bf Abstract}
\end{center}

\noindent
We find the Goldstino action descending from the N=1 Goldstone-Maxwell 
superfield action associated with the spontaneous partial supersymmetry 
breaking, N=2 to N=1, in superspace. The new Goldstino action has higher
(second-order) spacetime derivatives, while it can be most compactly described
as a solution to the simple recursive relation. Our action seems to be 
related to the standard (having only the first-order derivatives) 
Akulov-Volkov action for Goldstino via a field redefinition. 
\newpage

\section{Introduction}

Spontaneously broken global supersymmetry plays important role in brane-world 
models. D-branes generically break some part of supersymmetry that is, 
however, still {\it non-linearly} realised in the effective field theory in 
the D-brane worldvolume. The effective Lagrangians are highly constrained 
by the spontaneously broken supersymmetry while in some cases matter couplings
 can be determined exactly (see e.g., ref.~\cite{klein} for a recent review).

Spontaneously broken global supersymmetry is always accompanied by a 
(Nambu-Goldstone-type) fermionic (spin-1/2) particle called Goldstino 
\cite{wein}. The invariant Goldstino action (at least in its minimal form, as
an extension of Dirac action) is supposed to be fixed by the spontaneously 
broken supersymmetry up to field redefinition. Checking the independence of a 
given Goldstino action upon its origin (i.e. its {\it universality}) is often 
a highly non-trivial excercise in practice. In this Letter we give a new 
relevant example of this universality.

Our paper is organized as follows. In sect.~2 we review the standard 
{\it Akulov-Volkov} (AV) action \cite{av} for Goldstino, originating from the 
universal (coset) formalism of non-linear realizations. In sect.~3 we review
the N=1 supersymmetric {\it Bagger-Galperin} (BG) action \cite{bg}, 
originating from the spontaneous partial supersymmetry breaking, N=2 to N=1, 
in superspace. The new Goldstino action, descending from the BG action, is
derived in sect.~4. A connection to the AV action by field redefinition is
discussed in sect.~5. Our conclusions are summarized in sect.~6. We use the 
standard (2-component) notation for spinors in four (Minkowski) spacetime 
dimensions, as is given e.g., in ref.~\cite{wb}.

\section{Akulov-Volkov action}

The standard Goldstino action associated with the non-linearly realised 
supersymmetry is the so-called Akulov-Volkov (AV) action \cite{av}. A 
supersymmetry transformation is most naturally defined in superspace as a
shift of the superspace coordinates,
$$ \eqalign{
x^m ~\to~ & x'{}^m = x^m +i(\q\s^m\bar{\ve}-\ve\s^m\bar{\q})~,\cr
\q ~\to~ & \q'= \q + \ve~, \quad \bar{\q} ~\to~ \bar{\q}'=\bar{\q} +\bar{\ve}
~. \cr}\eqno(2.1)$$
The AV Goldstino spinor  $\l$ of broken N=1 supersymmetry can be viewed as a
superspace hypersurface defined by \cite{ik}
$$ \q = -\k \l(x)~, \eqno(2.2)$$
where the dimensional coupling constant $\k$ of mass dimension $-2$ has been
introduced. The coupling constant $\k$ determines the supersymmetry breaking
scale. Requiring the hypersurface (2.2) to be invariant under the 
transformations (2.1), i.e. $\q'(x)=\q(x')$, gives rise to the standard 
non-linear supersymmetry
transformation law \cite{av}
$$ \d_{\ve}\l=
\fracmm{1}{\k} \ve + i\k(\ve\s^m\bar{\l}-\l\s^m\bar{\ve})\pa_m\l~.
\eqno(2.3)$$
The non-linear transformations (2.3) represent the standard supersymmetry
algebra
$$  \[ \d_{\ve_1}, \d_{\ve_2} \] \l = -2i(\ve_1\s^m\bar{\ve}_2-\ve_2\s^m
\bar{\ve}_1)\pa_m\l~. \eqno(2.4)$$

To construct an invariant action one defines 
$$ \o\du{m}{n} =\d\du{m}{n} +i\k^2(\l\s^n\pa_m\bar{\l}-\pa_m\l\s^n\bar{\l})
\eqno(2.5)$$
so that $\det(\o)$ transforms as a density under the non-linear supersymmetry, 
$$ \d_{\ve}\det(\o)=-i\k\pa_m\left[ (\l\s^m\bar{\ve}-\ve\s^m\bar{\l})
\det(\o)\right]~.\eqno(2.6)$$
The invariant AV action reads \cite{av}
$$ S[\l] = -\fracmm{1}{2\k^2}\int d^4x\,\det\o =-\fracmm{1}{2\k^2}\int
d^4x -i\int d^4x\, \l\s^m\pa_m\bar{\l} +{\rm interaction~~terms}~~.
\eqno(2.7)$$
The AV Lagrangian has, therefore, the non-vanishing vacuum expectation 
value $-1/2\k^2$, while its leading term is given by the Dirac Lagrangian, 
as it should. The N=1
supersymmetry is entirely realised in terms of a {\it single} fermionic field 
$\l(x)$. It is possible to construct various supermultiplets of {\it linear} 
supersymmetry in terms of the Goldstino field of the  {\it non-linear} 
supersymmetry, order by order in $\k$, see e.g., refs.~\cite{ik,ro}. 
For a later use, we calculate the leading interaction terms in the AV 
Lagrangian, 
$$\eqalign{
 L_{\rm AV} = & -\fracmm{1}{2\k^2}\det{\o}=
-\fracmm{1}{2\k^2} -\fracmm{i}{2}(\l\s^m\pa_m\bar{\l}-
\pa_m\l\s^m\bar{\l})\cr
& -\fracmm{\k^2}{4}(\l\s^n\pa_m\bar{\l}-\pa_m\l\s^n\bar{\l})
(\l\s^m\pa_n\bar{\l}-\pa_n\l\s^m\bar{\l}) \cr
& +\fracmm{\k^2}{4}(\l\s^m\pa_m\bar{\l}-\pa_m\l\s^m\bar{\l})
(\l\s^n\pa_n\bar{\l}-\pa_n\l\s^n\bar{\l}) + {\cal O}(\k^4)~.\cr}\eqno(2.8)$$

\section{Bagger-Galperin action}

The D-brane worldvolume effective action necessarily contains the terms 
describing a selfinteracting abelian vector field $A_{\m}\,$. In the case of a
single `spacetime-filling' D3-brane, which is relevant here, those (low-energy)
 terms are given by the (1+3)-dimensional {\it Born-Infeld} (BI)
action \cite{bi,lei}
$$ S_{\rm BI}= \fracmm{1}{\k^2}\int d^4x\, \left( 1-\sqrt{-\det(\h_{mn}+
\k F_{mn})}\right)~, \eqno(3.1)$$
where $F_{mn}=\pa_{m}A_{n}-\pa_{n}A_{m}$ is the abelian field strength.
The photon field strength $F_{mn}$ can be extended to the N=1 chiral spinor
superfield strength $W_{\a}$ of an N=1 vector (Maxwell) supermultiplet as 
follows \cite{wb}:
$$ \eqalign{
W_{\a}(x,\q,\bar{\q}) ~=~ & \j_{\a}-i(\s^{mn}\q)_{\a}F_{mn} +\q_{\a}D
- i\q^2(\s^m\pa_{m}\bar{\j})_{\a} \cr
 & +i(\q\s^m\bar{\q})\pa_m\j_{\a} -\frac{1}{2}\q^2(\s^m\bar{\q})_{\a}
(i\pa_{m}D-\pa^nF_{mn})+\frac{1}{4}\q^2\bar{\q}^2\bo\j_{\a}\cr}\eqno(3.2)$$
that contains, in addition to $F_{mn}(x)$ its fermionic superpartner (photino)
$\j_{\a}(x)$ and the real auxiliary scalar $D(x)$. We use the notation
$\q^2=\q^{\a}\q_{\a}$ for chiral spinors and spinor covariant derivatives 
(see below), and similarly for the anti-chiral ones. The superfield $W_{\a}$ 
 satisfies the N=1 superspace Bianchi identities \cite{wb}
$$ \bar{D}_{\dt{\a}}W\low{\a}=0\quad {\rm and}\quad D^{\a}W_{\a}=
\bar{D}_{\dt{\a}}\bar{W}^{\dt{\a}}~~, \eqno(3.3)$$
where we have introduced the standard N=1 supercovariant derivatives in
superspace, 
$$ D_{\a} = \fracmm{\pa}{\pa\q^{\a}}+i(\s^m\bar{\q})_{\a}\pa_{m}~, \quad
\bar{D}_{\dt{\a}}=-\fracmm{\pa}{\pa\bar{\q}^{\dt{\a}}}
-i(\q\s^m)_{\dt{\a}}\pa_m~~, \eqno(3.4)$$
obeying the relations
$$ \{D_{\a}, D_{\b}\}=0\quad {\rm and}\quad \{D\low{\a},  \bar{D}_{\dt{\b}}\}
=-2i\s^m_{\a\dt{\b}}\pa_m~~.\eqno(3.5)$$

The manifestly N=1 supersymmetric generlization of the BI action (3.1) was
found by Bagger and Galperin \cite{bg} in the form~\footnote{See 
ref.~\cite{ket} for the manifestly N=2 supersymmetric generalization of the
BI action.}
$$ S_{\rm BG} = \int d^4x\,L_{\rm BG} = 
\int d^4x\left( \int d^2\q\, X + {\rm h.c.}\right)~, 
\eqno(3.6)$$
whose N=1 chiral superfield Lagrangian $X$ is determined by the recursive
formula \cite{bg}
$$ X = \fracmm{\fracmm{1}{4}W^2}{1+ \fracmm{\k^2}{4}\bar{D}^2\bar{X}}~~.
\eqno(3.7)$$
It follows from eqs.~(3.6) and (3.7) that 
$$L_{\rm BG} = \left[ \fracmm{1}{4}\int d^2\q\, W^2 +{\rm h.c.}\right]
+\fracmm{\k^2}{8} \int d^2\q d^2\bar{\q}\, W^2\bar{W}^2 + {\cal O}(\k^4)~, 
\eqno(3.8)$$
whose leading terms describe the N=1 supersymmetric Maxwell Lagrangian, as they
should. The exact solution to the non-linear constraint (3.7) is given by
\cite{bg}
$$ X = \fracmm{1}{4}W^2-\fracmm{\k^2}{32}\bar{D}^2\left[ \fracmm{
W^2\bar{W}^2}{1-\fracmm{1}{2}A+\sqrt{1+\fracmm{1}{4}B^2-A}}\right]~, 
\eqno(3.9)$$
where
$$\eqalign{
 A = & -\fracmm{\k^2}{8}(D^2W^2+\bar{D}^2\bar{W}^2)~, \cr
 B = & -\fracmm{\k^2}{8}(D^2W^2-\bar{D}^2\bar{W}^2)~. \cr}\eqno(3.10)$$
As regards natural {\it non-abelian} N=1 and N=2 supersymmetric 
extensions of the BI action (3.1),  see ref.~\cite{ket2}.

By construction the N=1 BI action (3.6) has manifest (linearly realised) N=1
supersymmetry. On the top of it there is another (non-linear) supersymmetry, 
whose transformation law is given by \cite{bg}
$$\d_{\h}W_{\a} = \left( \fracmm{2}{\k}+\fracmm{\k}{2}\bar{D}^2\bar{X}\right)
  \h_{\a} +2i\k(\s^m\bar{\h})_{\a}\pa_m X~~.\eqno(3.11)$$
The invariance of the action (3.6) under the transformation (3.11) is highly
non-trivial, given the fact that $X$ is a complicated function of $W$ and 
$\bar{W}$,  see eq.~(3.9). In addition, eq.~(3.11) is fully consistent with
 the N=1 superfield Bianchi identities (3.3). The invariance of the N=1 
superfield BI action (3.6) under the transformation (3.11) is technically a
consequence of the fact that \cite{bg}
$$ \d_{\h} X = \fracmm{1}{\k}W^{\a}\h_{\a}~~.\eqno(3.12)$$

The bosonic BI action (3.1) is recovered from eq.~(3.6) after integrating over
$\q$'s and setting $\j_{\a}=D=0$. We are going to consider now the purely
{\it fermionic} terms in the N=1 BI action by setting
$$ F_{mn} = D = 0~.\eqno(3.13)$$
Of course, the truncation (3.13) of the action (3.6) explicitly breaks the
linear N=1 supersymmetry. However, it is still consistent with the second
{\it non-linearly} realised supersymmetry (3.11) because that leaves the
constraint (3.13) to be invariant, i.e.
$$ \d_{\h} \left. (D_{\a}W_{\b})\right|_{F_{mn}=D=0}=0~~, \eqno(3.14)$$
where  $\left.\right|$ stands for the first (leading) component of a 
superfield or an operator. Being subject to extra constraint (3.13), 
the N=1 supersymmetric BG action (3.6) thus gives rise to a new Goldstino
action in terms of the Goldstino spinor field $\j_{\a}(x)$ alone. To the best 
of our knowledge, the fermionic terms in the BG action were not investigated 
in the literature yet. 

\section{New Goldstino action}

The purpose of this section is to derive the new Goldstino action $S[\j]$
descending from the BG action (3.6) after imposing the constraints (3.13).

The chiral superfield $X$ can be expanded in its field components 
$(\f, \c^{\a}, F)$ as follows:
$$ \eqalign{
X = & \f +\q\c + \q^2 F +i(\q\s^m\bar{\q})\pa_m\f \cr
 & -\fracmm{i}{2}\q^2(\pa_m\c\s^m\bar{\q}) +\fracmm{1}{4}\q^2\bar{\q}^2\bo\f~, 
\cr}\eqno(4.1)$$
so that we have
$$ \left. X\right|=\f(x), \quad \int d^2\q\, X =-\fracmm{1}{4}\left.D^2X\right|
= F(x)~.\eqno(4.2)$$
The constraints (3.13) imply 
$$ \c_{\a}=D_{\a}\!\left.X\right|=0~.\eqno(4.3)$$
As is clear from eqs.~(3.6) and (4.2), we have
$$ S[\j]= \int d^4x\, L = \int d^4x\, F +{\rm h.c.}\eqno(4.4)$$

It is not difficult to derive the recursive relation on the field $F$ and
its complex conjugate $\bar{F}$ from the recursive relation (3.7) on the
superfields $X$ and its complex conjugate $\bar{X}$ by using eqs.~(3.13) and 
(4.1), as well as the identities
$$\eqalign{
D^2W\low{\a} =~ & 4i\s^m_{\a\dt{\b}}\pa_m\bar{W}^{\dt{\b}}~,\cr
D^2\bar{D}^2\bar{W}^2 =~ & 16\bo\bar{W}^2~,\cr}\eqno(4.5)$$
and
$$ \eqalign{
D_{\a}X = & -\fracmm{\fracmm{1}{2}W^{\b}D_{\a}W_{\b}}{1+\fracmm{\k^2}{4}
\bar{D}^2\bar{X}} + \fracmm{\fracmm{i\k^2}{4}W^2\s^m_{\a\dt{\b}}
\bar{D}^{\dt{\b}}\pa_m\bar{X}}{\left(1+\fracmm{\k^2}{4}\bar{D}^2\bar{X}
\right)^2}~~,\cr
D^2X= & \fracmm{2iW\s^n\pa_n\bar{W}}{1+\fracmm{\k^2}{4}\bar{D}^2\bar{X}}
- \fracmm{\k^2W^2\bo\bar{X}}{(1+\fracmm{\k^2}{4}\bar{D}^2\bar{X})^2}
~~.\cr}\eqno(4.6)$$
As a result, we arrive at a non-linear constraint on $F(\j,\bar{\j})$ 
in the form
$$ -4F(1-\k^2\bar{F})^2=2i(\j\s^n\pa_n\bar{\j})(1-\k^2\bar{F})-
\fracmm{1}{4}\k^2\j^2\bo\left[ \fracmm{\bar{\j}^2}{1-\k^2F}\right]
\eqno(4.7)$$
and its complex conjugate. As will be shown in Sect.~5, despite of the 
apparent presence of higher derivatives in eq.~(4.7), this equation does not 
imply the equations of motion. Instead, it should be considered as the 
off-shell recursive relation on the Lagrangian $L(\j,\bar{\j})=2{\rm Re}(F)$. 
Equation (4.7) fully determines $F$ and, hence, the action (4.4) in terms of 
$\j$, $\bar{\j}$ and their spacetime derivatives. 

The leading term in $L$ is given by the Dirac Lagrangian, as it should, 
$$F_0\equiv \left.F\right|_{\k^2=0}=-\fracmm{i}{2}\j\s^m\pa_m\bar{\j}\equiv
\fracmm{1}{2} L_0~~.\eqno(4.8)$$
The exact (i.e. to all orders in $\k^2$) solution to the non-linear constraint
(4.7) descends from the superfield solution (3.9). A straightforward (albeit
 tedious) calculation yields
$$ \eqalign{
F  & =  \fracmm{1}{2}L_0 + \fracmm{\fracmm{\k^2}{8}\left(4\abs{L_0}^2+\j^2\bo
\bar{\j}^2\right)}{1-\fracmm{1}{2}A_0 +\sqrt{1+\fracmm{1}{4}B^2_0-A_0}} \cr
& + \fracmm{\fracmm{\k^4}{16}\bar{L}_0\j^2\bo\bar{\j}^2}{\left(
1-\fracmm{1}{2}A_0 +\sqrt{1+\fracmm{1}{4}B^2_0-A_0}\right)^2}\left[
1 + \fracmm{1+\fracmm{1}{2}B_0}{\sqrt{1+\fracmm{1}{4}B^2_0-A_0}}\right] \cr
& + \fracmm{ \fracmm{\k^4}{16}\left\{ \j^2\bar{\j}^2\bo L_0 +
2\j^2\pa^m\bar{\j}^2\pa_mL_0 + L_0\bar{\j}^2\bo\j^2 \right\}} {\left(
1-\fracmm{1}{2}A_0 +\sqrt{1+\fracmm{1}{4}B^2_0-A_0}\right)^2}\left[
1 + \fracmm{1-\fracmm{1}{2}B_0}{\sqrt{1+\fracmm{1}{4}B^2_0-A_0}}\right] \cr
& + \fracmm{\fracmm{\k^6}{64}\j^2\bar{\j}^2 \bo\j^2\bo\bar{\j}^2} {\left(
1-\fracmm{1}{2}A_0 +\sqrt{1+\fracmm{1}{4}B^2_0-A_0}\right)^3}
\left[ 1+ \fracmm{1+\fracmm{1}{2}B_0}{\sqrt{1+\fracmm{1}{4}B^2_0-A_0}}\right]
\left[ 1+ \fracmm{1-\fracmm{1}{2}B_0}{\sqrt{1+\fracmm{1}{4}B^2_0-A_0}}\right]
\cr 
& + \fracmm{\fracmm{\k^6}{128}\j^2\bar{\j}^2 \bo\j^2\bo\bar{\j}^2} {\left(
1-\fracmm{1}{2}A_0 +\sqrt{1+\fracmm{1}{4}B^2_0-A_0}\right)^2}
\fracmm{1}{\sqrt{1+\fracmm{1}{4}B^2_0-A_0}}
\left[ 1+ \fracmm{1-\fracmm{1}{4}B^2_0}{1+\fracmm{1}{4}B^2_0-A_0}\right]
\cr
& - \fracmm{\fracmm{\k^6}{32}\j^2\bar{\j}^2 \pa^mL_0\pa_mL_0}{\left(
1-\fracmm{1}{2}A_0 +\sqrt{1+\fracmm{1}{4}B^2_0-A_0}\right)^2}
\fracmm{1}{\sqrt{1+\fracmm{1}{4}B^2_0-A_0}}\left[
1 -\left( \fracmm{1-\fracmm{1}{2}B_0}{\sqrt{1+\fracmm{1}{4}B^2_0-A_0}}\right)^2
\right] \cr
& + \fracmm{\fracmm{\k^6}{16}\j^2\bar{\j}^2 \pa^mL_0\pa_mL_0} {\left(
1-\fracmm{1}{2}A_0 +\sqrt{1+\fracmm{1}{4}B^2_0-A_0}\right)^3}\left[
1 + \fracmm{1-\fracmm{1}{2}B_0}{\sqrt{1+\fracmm{1}{4}B^2_0-A_0}}\right]^2~, 
\cr}\eqno(4.9)$$
where we have used eq.~(3.10) and the notation
$$ A_0= \left.A\right|=2\k^2{\rm Re}(L_0)~, \quad
 B_0= \left.B\right|=2\k^2i{\rm Im}(L_0)~. \eqno(4.10)$$

The most important property of the action (4.4) is its invariance under the
following (rigid) non-linear supersymmetry transformations,
$$\d_{\h}\j_{\a}=\left(\fracmm{2}{\k}-2\k\bar{F}\right)\h_{\a}
+\fracmm{i}{2}\k(\s^m\bar{\h})_{\a}\pa_m\left(\fracmm{\j^2}{1-\k^2\bar{F}}
\right)~, \eqno(4.11)$$
descending from eq.~(3.11). In particular, we have $\VEV{0|\d_{\h}\j_{\a}|0}=
\fracmm{2}{\k}\h^{\a}\neq 0$, i.e. the vacuum is not supersymmertric. This
fact allows us to call $\j_{\a}(x)$ the Goldstino field, so that we can 
identify the action (4.4) as the Goldstino action associated with the 
spontaneous N=1 supersymmetry breaking. 

Despite of the apparent presence of many square roots in the Lagrangian (4.9), 
it is, in fact, {\it polynomial} in $\j$ and $\bar{\j}$ due to their 
anticommutativity that implies the nilponency conditions
$$ \j^3=\bar{\j}^3=L^3_0=\bar{L}_0^3=A_0^5=B_0^5=0~~.\eqno(4.12)$$

\section{Relation between the AV and BG actions}

The new Goldstino action $S[\j]$ descending from the BG action (see sect.~4) 
seems to be very different from the standard AV action $S[\l]$ (see sect.~2).
For example, the action $S[\l]$ has only the first-order spacetime 
derivatives of $\l$,  whereas the action $S[\j]$  has the second order 
derivatives of $\j$ also.

Both actions can be used as the Goldstino action because they are invariant 
under the corresponding non-linear supersymmetry transformations having 
inhomogeneous shifts proportional to the anticommuting spinor parameter $\h$.

Whatever the reason for a spontaneous global symmetry breakdown, the broken 
symmetry is supposed to fix the invariant (minimal) Goldstone action up to
field redefinition. The anticipated universality of the Goldstino action gives
us a good reason to suspect that the actions $S[\l]$ and  $S[\j]$ may be 
equivalent up to a field redefinition. We are now going to check this 
conjecture up to the order $\k^2$.

The difference between the actions $S[\l]$ and  $S[\j]$ can be thought of as
the direct consequence of the difference between the corresponding 
supersymmetry transformation laws in eqs.~(2.3) and (4.11), respectively.
The transformation law (2.3) originated from the general (coset) approach to
non-linear realizations of N=1 supersymmetry, whereas the BG-type 
transformation law (4.11) appeared out of the context of partial spontaneous 
supersymmetry breaking, N=2 to N=1, when first making manifest the unbroken 
N=1 supersymmetry of the corresponding Goldstone-Maxwell action. Hence, the 
field redefinition in question can be found by applying the known relation 
between linear and non-linear realizations of supersymmetry \cite{ik,iva}. 
 
The structure of the supersymmetry transformations (3.11) and (3.12) implies
that the fields $\f(x)=\left.X\right|$ and $\j_{\a}(x)=\left.W_{\a}\right|$ 
are, in fact, the components of an N=1 {\it chiral} superfield with respect 
to the second N=1 supersymmetry, 
$$\eqalign{
X(x, \z, \bar{\z})= & ~\f +\fracmm{1}{\k}\j\z + \left(\fracmm{1}{\k^2}-\bar{F}
\right)\z^2 +i(\z\s^m\bar{\z})\pa_m\f \cr
& - \fracmm{i}{2\k}\z^2(\pa_m\j\s^m\bar{\z}) +\fracmm{1}{4}\z^2\bar{\z}^2\bo
\f~, \cr}\eqno(5.1)$$
since eqs.~(3.11) and (3.12) then follow from eq.~(2.1) in N=1 superspace
$(x, \z, \bar{\z})$~, 
$$ \d x^m=i(\z\s^m\bar{\h}-\h\s^m\bar{\z})~, \quad \d\z=\h~, \quad
\d\bar{\z}=\bar{\h}~~.\eqno(5.2)$$

The procedure of passing to the standard non-linear realization of 
supersymmetry, in the case of partial spontaneous supersymmetry breaking N=2
to N=1, was formulated in ref.~\cite{iva}. It equally applies to our case of 
spontaneous supersymmetry breaking N=1 to N=0. The idea is to consider the
{\it finite} $\h$-transformations of the fields $\f$ and $\j_{\a}\,$, which are
generated by the infinitesimal supersymmetry transformations (3.11) and
(3.12), 
$$ \left( \begin{array}{c} \tilde{\f}(\h) \\ \tilde{\j}_{\a}(\h) 
\end{array}
\right) =  \left( 1 +\d_{\h} +\fracmm{1}{2!}\d^2_{\h} 
+\fracmm{1}{3!}\d^3_{\h} +\fracmm{1}{4!}\d^4_{\h}\right)
\left( \begin{array}{c} \f \\ \j_{\a} \end{array}\right)~~,\eqno(5.3)$$
and then replace the anticommuting parameters $(\h,\bar{\h})$ by the AV 
fermions, by using the standard rule (2.2),~\footnote{We rescaled $\l$ by
the factor of $1/2$ for a later convenience.}
$$ \h_{\a}\to -\fracmm{\k}{2}\l_{\a}~, \quad 
\bar{\h}_{\dt{\a}}\to -\fracmm{\k}{2}\bar{\l}_{\dt{\a}}~.\eqno(5.4)$$
The composite fields $\tilde{\f}(\l)$ and  $\tilde{\j}_{\a}(\l)$ transform 
(non-linearly) homogeneously under the $\h$-transformations \cite{iva},
$$ \d_{\h}\left( \begin{array}{c} \tilde{\f}(\l) \\ \tilde{\j}_{\a}(\l) 
\end{array}\right) = \fracmm{i}{2} \left( \l^{\a}\bar{\h}^{\dt{\a}}-
\h^{\a}\bar{\l}^{\dt{\a}}\right)\pa_{\a\dt{\a}} \left( 
\begin{array}{c} \tilde{\f}(\l) \\ \tilde{\j}_{\a}(\l)\end{array}\right) ~~,
\eqno(5.5)$$
so that the constraints
$$  \tilde{\f}(\l) = \tilde{\j}_{\a}(\l)=0 \eqno(5.6)$$
are invariant under the non-linear supersymmetry. Equations (5.6) give us the
desired relation between the spinor fields $\j(x)$ and $\l(x)$ in
the closed (though rather implicit) form ({\sl cf.} refs.~\cite{ro,iva}): 
$$\eqalign{
0 & =  \f - \fracmm{\l^2}{4}\left(1-\k^2\bar{F}\right) 
- \fracmm{i\k^2}{4}\left(\l\s^m\bar{\l}\right)\pa_m\f \cr
& -\fracmm{i\k^2}{8}\l^2\left(\pa_m\j\s^m\bar{\l}\right)
+\fracmm{3\k^2}{64}\l^2\bar{\l}^2\bo\f \cr}\eqno(5.7)$$
and
$$\eqalign{
0 & =  \j_{\a} -\l_{\a}(1-\k^2\bar{F})-i\k^2(\s^m\bar{\l})_{\a}\pa_m\f 
 -\fracmm{i\k^2}{4}\left(\pa_m\j\s^m\bar{\l}\right)\l_{\a}
+ \fracmm{i\k^2}{4}(\s^m\bar{\l})_{\a}(\l\pa_m\j) \cr 
& + \fracmm{i\k^4}{8}\l^2(\s^m\bar{\l})_{\a}\pa_m\bar{F} 
 - \fracmm{\k^4}{8}\l_{\a}\bar{\l}^2\bo\f - \fracmm{\k^4}{64}\l^2\bar{\l}^2
\bo\j_{\a}~,\cr}\eqno(5.8)$$
where the field $F$ is given by eq.~(4.7) or (4.9).~\footnote{Equation (5.8) 
was used in deriving eq.~(5.7) from eq.~(5.3).} Equations (5.7) and 
(5.8) can be used to unambiguously calculate $\j$ as a function of $\l$ 
order-by-order in $\k^2$, after eliminating the auxiliary fields $\f$ and $F$.
The invariant (under the non-linear supersymmetry) relation between $\j$ and 
$\l$ in eq.~(5.8) is not universal because the Goldstino Lagrangian 
$L=2{\rm Re}(F)$ enters eq.~(5.8) as the auxiliary field, see eq.~(4.4).

When keeping only the terms of order $\k^0$ and $\k^2$, we find the leading
and subleading terms in the action (4.4) as follows:
$$\eqalign{
L = F +\bar{F}= & -\fracmm{i}{2}\j\s^n\pa_n\bar{\j}
+\fracmm{i}{2}\pa_n\j\s^n\bar{\j} 
+\fracmm{\k^2}{2}(\j\s^n\pa_n\bar{\j})(\pa_m\j\s^m\bar{\j}) \cr
 & \fracmm{\k^2}{16}\j^2\bo\bar{\j}^2 + \fracmm{\k^2}{16}\bar{\j}^2\bo\j^2
+ {\cal O}(\k^4)~.\cr}\eqno(5.9)$$
Having substituted eqs.~(5.7) and (5.8) into eq.~(5.9) up to the given order 
in $\k^2$, we recovered  eq.~(2.8) by using Fierzing identities and 
integration by parts. This pattern may persist to all orders in $\k^2$, though
we still cannot exclude a possible higher-order superinvariant, with at least
four spacetime derivatives, as a non-trivial difference between $S[\j(\l)]$ 
and $S[\l]$.

\section{Conclusion}

The non-linear realizations formalism gives rise to the AV action  $S[\l]$ for
Goldstino without higher derivatives. When one wants to assign Goldstino to a 
vector (Maxwell) supermultiplet (thus identifying Goldstino and photino), in 
the context of partial spontaneous supersymmetry breaking, one gets another 
Goldstino action $S[\j]$ that has higher (second-order) derivatives. The new 
action $S[\j]$ appears to be equivalent to the AV action $S[\l]$ up to a field
redefinition and integration by parts. By providing the field redefinition in 
question we verified the universality of Goldstino action up to the terms of 
order $\k^2$. The proposed on-shell equivalence of the actions  $S[\l]$ and  
$S[\j]$ is highly non-trivial because of the complicated relation between the 
Goldstino fields $\l$ and $\j$, described by eqs.~(4.9), (5.7) and (5.8).

\newpage

\section*{Acknowledgements}

The authors are grateful to Augusto Sagnotti, whose question led to this 
investigation, Jim Gates Jr. and Satoru Saito for useful discussions, and the
referee for his careful reading of our manuscript. One of the authors (S.V.K.)
would like to thank the Center for String and Particle Theory (CSPT) at the 
University of Maryland in College Park (USA) for kind hospitality during some 
part of this work.

\end{document}